# Radiotherapy using a laser proton accelerator

Masao Murakami,[a] Yoshio Hishikawa,[a] Satoshi Miyajima,[b] Yoshiko Okazaki,[b] Kenneth L. Sutherland,[c] Mitsuyuki Abe,[d] Sergei V. Bulanov,[e] Hiroyuki Daido,[e] Timur Zh. Esirkepov,[e] James Koga,[e] Mitsuru Yamagiwa,[e] Toshiki Tajima[e]

[a]*Department of Radiology, Hyogo Ion Beam Medical Center, 1-2-1, Kouto, Tatsuno, Hyogo 679-5165, Japan*

*Division of Medical Imaging & Ion Beam Therapy, Kobe University Graduate School of Medicine, 7-5-1, Kusunokicho, Chuo-ku, Kobe, Hyogo 650-0017, Japan*

*CREST, Japan Science and Technology Agency, 4-1-8, Honcho, Kawaguchi, Saitama 332-0012, Japan*

[b]*CREST, Japan Science and Technology Agency, 4-1-8, Honcho, Kawaguchi, Saitama 332-0012, Japan*

[c]*CREST, Japan Science and Technology Agency, 4-1-8, Honcho, Kawaguchi, Saitama 332-0012, Japan*

*School of Medicine and Health Sciences, Hokkaido University, North 15, West 7, Kita-ku, Sapporo 060-8638, Japan*

[d]*Department of Radiology, Hyogo Ion Beam Medical Center, 1-2-1, Kouto, Tatsuno, Hyogo 679-5165, Japan*

[e]*Photo-Medical Reseach Center, Japan Atomic Energy Agency, 8-1 Umemidai, Kizugawa, Kyoto 619-0215, Japan*

*CREST, Japan Science and Technology Agency, 4-1-8, Honcho, Kawaguchi, Saitama 332-0012, Japan*

**Abstract.** Laser acceleration promises innovation in particle beam therapy of cancer where an ultra-compact accelerator system for cancer beam therapy can become affordable to a broad range of patients. This is not feasible without the introduction of a technology that is radically different from the conventional accelerator-based approach. Because of its compactness and other novel characteristics, the laser acceleration method provides many enhanced capabilities for the radiation oncologist. First, it reduces the overall system size and weight by more than one order of magnitude. This is due not only to the compactness of the accelerator component, but also to reduced radiation shielding requirements, fewer proton magnets, and the compactness of the gantry. In addition to this ultra-compact size, the characteristics of the particle beams (protons) make them suitable for a class of therapy that might not be possible with the conventional accelerator, such as the ease for changing pulse intensity, the focus spread, the pinpointedness, and the dose delivery in general. Some new methods of therapy may be derived from these characteristics. A compact, uncluttered system allows a PET device to be located in the vicinity of the patient in concert with the compact gantry. The radiation oncologist may be able to irradiate a localized tumor by scanning with a pencil-like particle beam while ascertaining the actual dosage in the patient with an improved in-beam PET verification of auto-radioactivation induced by the beam therapy. This should yield an unprecedented flexibility in the feedback radiotherapy by the radiation oncologist. Assisted by this, irradiation scanning based on the minute beam structure of laser accelerated particle beams well conforms to the small size of early stage tumors. In order to improve advance experimental achievement, we introduce a series of innovations intended to enhance the energy and quality of accelerated ions with the laser acceleration method. This technology will drive the development of a new class of beam therapy, which is highly localized, specific and has been termed biologically conformal radiation therapy (BCRT). In a clinical point of view, this compact and sophisticated laser accelerated proton radiotherapy at lower energies near 60-80MeV can be applicable to the disease lying a few centimeter under the skin, such as eye tumors, age-related macular degeneration (ARMD), laryngeal cancer, nasal or paranasal tumor, breast cancer, and so on. Laser accelerated radiotherapy has a unique niche in a current world of high energy accelerator using synchrotron or cyclotron.

# INTRODUCTION

Although the technique for targeting X- or γ-rays for tumor treatment volumes has greatly improved in recent years, the radiation exposure to normal tissues around the tumor is still significant and limits the maximum dose available for sterilizing tumor cells. Particle beams, including protons and carbon-ion beams, provide a way to resolve this problem, because damage to normal tissue around the tumor is minimized due the effects of the particles' Bragg peak.[1]

According to a worldwide survey report,[2] as of July 2004 there were over 48,000 patients (42,700 by proton, 1,100 by pions and 4,500 by ions) treated by particle ion beams at 25 institutes. Usually, particle beams are accelerated by a synchrotron or cyclotron facility. These conventional accelerators are based on radio-frequency (RF) technology. In addition, many bending magnets are required for ion beam transport to the procedure rooms and for providing conformal irradiation of the tumor target (the gantry systems). These must be accompanied by the adequate radiation shielding. Laser acceleration of charged particles,[3] however, may provide a viable alternative technology for hadron therapy.[4,5,6,7,8,9,10]

Since the first high power laser proton acceleration experiments[11,12,13] in 2000 there has been world-wide experimental research[14,15,16] to ascertain and improve the laser acceleration of ions, as well as more theoretical and computer simulations (see review articles[17,18,19] and literature quoted in). Various aspects of radiotherapy using laser accelerated ions have been previously addressed in the above cited papers. In this paper we aim to expand on some of these ideas. We address real time verification methods which are enabled by laser accelerated protons. We use previously presented multi-parametric studies to find an optimal scheme for acceleration of ions with double layer targets for increasing the maximum energy and efficiency for a given laser pulse energy. Although the basic mechanisms are quite well known from theory, simulations, and experiments, medical applications require new mechanisms of acceleration. In this paper we propose a new acceleration mechanism, adiabatic acceleration. The present paper is organized as follows: (1) as an important background for this paper, we will review the already known and newly recognized distinctive features and limitations of present laser driven proton accelerators, (2) examine methods for optimization including energy maximization and adiabatic acceleration, and (3) propose specifications of the laser accelerator for producing the required quality hadron beams, and examine the clinical availability and innovative potential for making these unique treatments feasible by this enabling technology. As a result we shall show a future vista of novel radiotherapy based on the laser accelerator approach.

## CHRACTERISTICS OF LASER DRIVEN PARTICLE THERAPY

### General

A laser ion accelerator is expected to be compact, simple, and low cost. Some of the reasons for this are shown in Table 1. This is based on several factors: We can compactify the system so that it can be accommodated in existing hospital infrastructures. The gantry is small, containing no large magnets (unlike charged particle beams, no magnets are necessary for laser photon transport) and radiation shielding is only local (instead of shielding an entire accelerator section, as in the conventional case). The accelerator itself is tiny (relative to conventional ones) where the most expensive element in our machine is the laser. Because lasers are rapidly getting cheaper and more powerful, it is anticipated that they will significantly contribute to the widespread use of proton therapy throughout the world. The irradiation system of a laser ion accelerator is a multi-terawatt system with a final laser pulse compressor, an ion beam generation chamber, a separation magnet system and a patient positioning system. These features are remarkable in comparison with synchrotron or cyclotron accelerators.[20] These days compact, few hundred TW lasers that can easily fit into typical hospital radiotherapy treatment rooms are commercially available. These innovative features amount to a proton therapy device that is comparable in size to currently employed linac X-ray machines in many typical hospitals. This allows one to reduce the cost of the infrastructure of a proton therapy facility. A typical experimental setup of laser acceleration of ions looks like that in Fig.1, being of miniature size.[21] In the future the real device will be an order of magnitude smaller than those laser systems and interaction chambers presently intended for fundamental science studies.

**TABLE 1.** Characteristics of a laser driven particle therapy machine

| Parameter | Conventional accelerator | Laser driven accelerator |
|---|---|---|
| Beam transport | Large magnetic system required to bend charge particles<br>Radiation shielding required around the bending corner | Laser beam can be transported and bended by mirrors (small device).<br>Radiation shielding not required around the bending corner |
| Accelerator | Large magnetic system required to accelerate charge particles<br>Total system is large and overall radiation shielding required | Only target part should be irradiated.<br>Compactification of the system possible |
| Irradiation system | The three to five meter magnet system required to bend charged particles 90-135 degrees with gantry | Small size magnet to bend 10-30 degrees (one tenth of conventional one in size) |
| Gantry | Large size (100-250 tons)<br>Diagnostic system cannot be set near patient | Small size (1-10 tons)<br>Diagnostic system can be combined with gantry. |
| Scanning | Mainly long pulse injection<br>Recently started | Short pulses superimposed<br>Main method |
| Technology | Matured (started since 1930s) | Nascent (started since 1990s)<br>Many elements to be developed |
| Others |  | Protective goggle required |

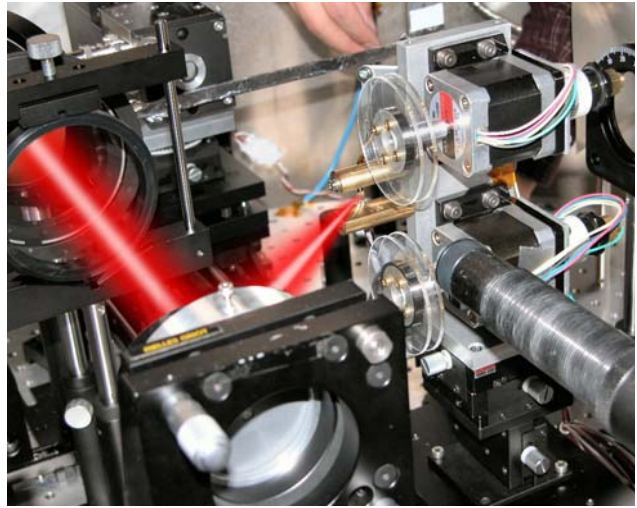

**FIGURE 1.** Experimental setup of repetitive laser-driven accelerator, which can be hand palm size. The actual medical device needs some energy separator and radiation shield etc., which amounts to a physical extent of order 1m according to Ref.[22] (see Fig.2). The laser pulse length, intensity, direction, spot position etc. are easily controlled so that the resultant proton beams may be produced to the desired specification.

## Compactness of the System

### Beam transport

The beam transport system of a laser ion accelerator is much different from that of conventional ion accelerators. Unlike the latter, ion acceleration would take place in the gantry (or certainly in the patient's room) rather than at a distant site. There would be no ion injector. There would be no large bulky transport magnets. Whereas charged particles such as protons and carbon ions are bent by magnets, laser light is redirected in transport by mirrors. The final mirror (or mirrors) would focus the laser onto a target that is itself an accelerator of the size comparable to the palm of a hand or at most an arm.

### Irradiation system and gantry

Another key advantage of laser accelerators is that they can be naturally incorporated into the transport system, i.e. they do not involve a large number of magnets.[5] The generated proton beam would be transported, bent, and focused onto the patient through a system of magnets and shields that separate unwanted components of radiation, including protons of undesired energies. Such an example has already been suggested.[5, 22] One of the reasons for the need for the large number of magnets in the gantry is that in conventional accelerators protons of 200 MeV need to be bent by magnets typically by 135 degrees. By contrast laser accelerated protons are generated much closer to the patient so the typical bending angle would be a mere 10 or 10's of degress.[22]

*Acceleration gradient*

In the case of a laser accelerator, the acceleration gradient is several orders of magnitude greater (of order ~100 GeV/m) than that of typical conventional RF technology (which is of order 10 to 100 MeV/m). This raises expectations for providing much more compact therapy devices. The limit for conventional accelerators is due to rf cavity breakdown.[23]

## Characteristics of laser-driven particle beams

*Laser energy delivery and therapy of small early tumors*

As diagnostic methods improve over the next decade (e.g., molecular imaging, magnetic resonance imaging, magnetic resonance spectroscopy and other advanced sensitive detection techniques or functional imaging), earlier detection of smaller tumors may become a reality. This not only helps to avoid metastasis, but also facilitates flexibility and ease of cure. We envision therapy of millimeter size tumors rather than centimeter size in the future. The conventional accelerator typically delivers $10^{10}$ particles per pulse. Many laser experiments that have been carried out so far can yield about $10^9$ particles per pulse (however, parameters widely vary and it is not so meaningful to mention just one parameter).

It is important to note that (1) in terms of energetics there is an ample margin for a laser accelerator to meet the desired fluence (or number) of protons, while (2) with regard to realizing this new technology that is radically different from conventional rf accelerators (synchrotron or cyclotron), one needs a series of innovations that match the requirements of medical applications. In many analytical[9] and simulation[24] studies, intensities of $10^{20-21}$ W/cm$^2$ are necessary to generate ions over 100 MeV. This suggests that we need laser irradiation of peak power 100 TW or more (if focused onto an area of $10^{-3}$ cm x $10^{-3}$ cm or smaller). A 100 TW laser pulse of duration 100 fsec contains 10 J of energy. Suppose that this laser operates at a 10Hz repetition rate and we irradiate for 100 seconds (of course, the rate can be changed to conform with the treatment program, but here we assume only one fraction) to deliver a 70Gy dose to an early stage tumor of mass, 1gram. The total needed proton energy deposited in the tumor is 0.07J which amounts to 0.07mJ per laser pulse. Because the single laser pulse energy under consideration is 10J, we only need to transfer a small fraction ($10^{-5}$) of the laser pulse energy into proton energies of therapeutic relevance. Some of the current experiments typically deliver 1% of laser pulse energy to protons, though again this efficiency varies from experiment to experiment. On the other hand, as we shall discuss later, if more efficient adiabatic acceleration is realized, the energy fraction converted from the laser to proton beam (energy conversion efficiency) is significantly greater, which means that we can consider much lower power laser systems which can be easier to develop and more compact.

*Flexible combination of therapy beams and diagnostics*

It is known that in-beam positron emission tomography (PET) is the only method for non-invasive in situ monitoring for hadron therapy.[25] The smaller gantry affords flexibility in installing in-beam PET. In addition development in precise irradiation and diagnostic imaging will make it possible to develop patient specific irradiation therapy which is termed biologically conformal radiation therapy (BCRT).[26] This will result in scanning therapy with increased accuracy with reduced dosage.

*Precise Irradiation*

To achieve precise irradiation of irregularly shaped tumor targets via beam scanning irradiation, studies have shown the following requirements for medical accelerators:[27,28] a) beam size control, b) intensity control, c) fast beam cut-off, and d) time-structure control. Although these requirements can be achieved using conventional accelerators,[28] it can be much easier to achieve them using laser accelerated particle beams.

Concerning a), the laser ion accelerator has the advantage of generating narrow proton beams. From several experimental or simulation data, narrow proton beams of 50-200 micron diameters[29] can be available. Consequently, a variety of beam sizes built up from narrow beams and small spots via beam scanning.

Concerning b), the proton beam intensity can be easily controlled by laser energy for a fixed laser pulse duration and spot size. The output proton beam energy and flux would change according to the modulation of the laser intensity. Changing the energy of the proton beams can be done more quickly than in current accelerators because the laser intensity can be changed easily and quickly by changing the laser energy. These features make it possible to treat minute shallow targets precisely using the spot scanning technique.[30] These features will become more attractive and relevant as the early detection and diagnosis of cancer are expected to improve over the next generation.

When dynamic beam modulation of spot scanning is used, further dose concentration to the tumor than in the static case is anticipated. A tumor is divided into individual voxels, and the laser pulse intensity of each voxel is modulated corresponding to the required proton beam energy. The variation of laser intensity from one voxel to another is rapid and the change of the proton energy is correspondingly rapid. So, laser intensity modulated proton therapy (IMPT) driven by laser is a promising method for spot scanning.[31] We are aiming at highly accurate particle radiotherapy.

Concerning c) and d), another unique feature of laser accelerated protons is the short pulse duration (typically picoseocnds instead of the hundreds of nanoseconds typical of a conventional accelerator). Because such short proton pulses can clearly be controlled by turning the laser pulses on or off, extremely fast pulse cut-off and time-structure control are possible in contrast to conventional accelerators. This is directly attributed to the picosecond temporal structure of the laser produced proton beams.

*Precise real time verification and image guided proton radiotherapy*

An additional requirement for precise irradiation is on-line or real time diagnostic imaging. The integration of diagnostics with the therapy allowing image guided therapy in intensity-modulated radiotherapy (called as tomotherapy) has already been recognized as highly desirable.[32] We previously reported that a PET camera can detect positron emitters, consisting mainly of $^{11}$C, $^{13}$N and $^{15}$O along the proton beam trajectory.[33] At HIBMC the PET examination room is situated near the treatment room. The PET image is clinically very useful for the examination of patients, because we can easily verify the extent, angle, and the volume irradiated by proton beams. A laser driven proton accelerator will be sufficiently compact to include an in-beam PET camera into the irradiation system (Fig. 2) that allows us to improve the usage of the in-beam diagnostic PET operation with the scanning radiotherapy treatment.[34,35,25] With current in-beam PET scanning, the counting statistics are low due to low detection efficiency over limited angles and random coincidence effects due to secondary gamma photon emission.[36,37] These gamma photons contaminate PET signals during beam irradiation, rendering the noise level of gamma radiation high compared with the positron annihilation gamma signal due to auto-radioactivation by proton irradiation. Suppression of this noise is achieved by placing time windows on the data (gating).[37] Compared to conventional accelerator systems, laser accelerated short particle beams enable for much more accurate specification and positioning of the time windows as well as a reduction of the integrated time over which this noise due to secondary gamma emission occurs. In addition due to the compact size of laser accelerators there are fewer constraints on the placement of a closed-ring PET detector, which has better resolution than a dual-head PET detector.[38] Consequently, improved real time verification of auto-activation will be available with higher counting statistics due to the larger possible detection solid angles and higher noise suppression. This is expected to clarify the relationship between the intensity of proton-induced activities and tissue components. Ultimately, we aim to control the prescribed radiation dosage in real time measuring positron emitters generated immediately after irradiation. Real time verification of proton dose allows precise and safe treatment of small tumors such as eye melanoma. For larger invasive tumors such as H&N tumors, we can apply the technique to improvement in local control by increasing the intensity to radio-resistant part in the tumor.

*Enhanced therapy accuracy using shorter lived positron emitters*

It is recognized that by having an in-beam PET detector that shorter lived positron emitters can be used for imaging.[39] Figure 3a illustrates proton irradiation of water, simulated by the Monte Carlo Particle and Heavy Ion Transport code System, PHITS [40] in the case where a proton pencil beam of energy 200 MeV with a 10 mm diameter. PHITS is used in various fields, such as radiation shielding of accelerator facilities, radiotherapy, and space technology. Nuclear reactions are clearly seen along the proton tracks. Figure 3b (3c) shows the two-dimensional distribution of the residual nuclei, $^{15}$O ($^{11}$C) in water determined by PHITS. These residual nuclei have a short half-life of 2 (20) minutes. Positrons are created during $\beta^+$-decay they are subsequently annihilated in encounters with electrons. A pair of gamma photons results from the electron-positron annililation process. Comparison of the figures 3b and 3c shows that the number of $^{15}$O residual nuclei is far greater than the $^{11}$C residual nuclei. This is mainly due to the larger production cross section.[i] This illustrates that, with in-beam PET, using $^{15}$O as the more abundant positron emitter with the shorter half life can enhance PET imaging.

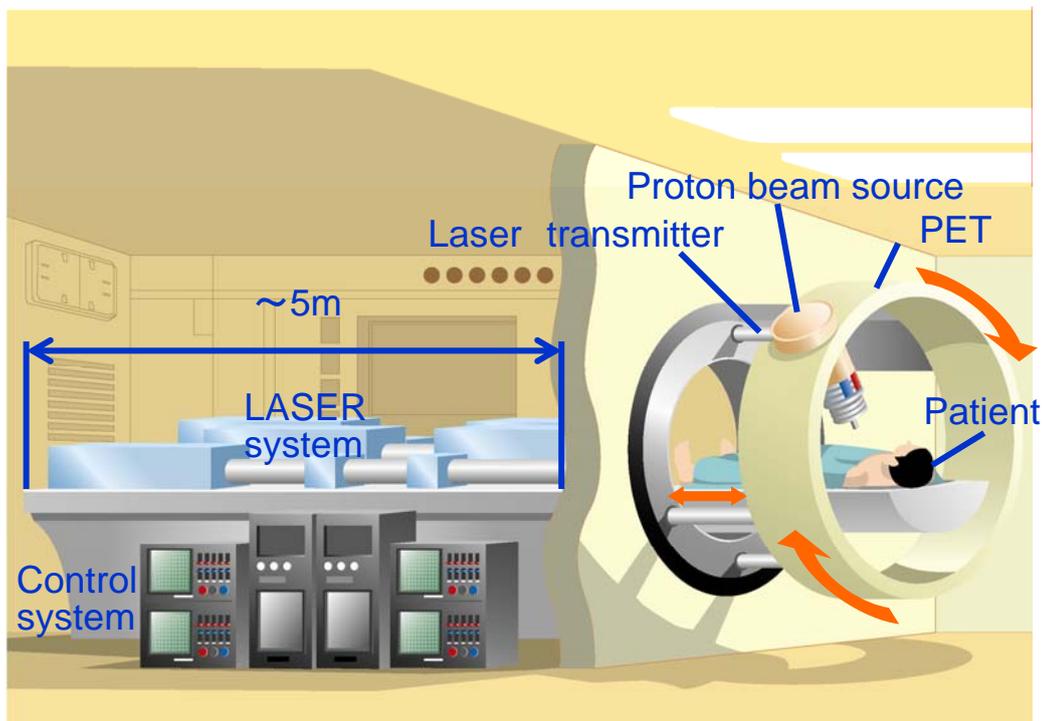

**FIGURE 2.** Conceptual laser accelerated proton therapy instrument. Compact and flexible radiotherapy and dose verification synergy due to the innovation of the laser proton acceleration combined with an in-situ real-time PET camera. The compactness of the laser acceleration introduces significant flexibility in operating real-time PET verification, which allows the radiation oncologist to render far more affordable and accurate cancer therapy to the patient.

## Limitations

To date, the maximum experimentally measured laser-accelerated proton energy has been limited to several tens of MeV.[17][18][19][42] These proton energies can penetrate to depths of several centimeters in the human body. With proton energies limited to 50 MeV, only 4 of 778 patients (0.5%) were treated between May 2003 and Oct 2005 at HIBMC. Three of the four were H&N cancer where two were nasal cavity malignant melanoma and one was an orbital tumor. The remaining one of the four was lung cancer near the chest wall. The proton ranges were 34mm, 45mm, 48mm, and 38mm, respectively.

Laser-accelerated proton beams are typically emitted with measurable diverging angles. However, the beam still originates from a tiny laser focus so that the transverse emittance is quite small (typically one order magnitude smaller than of conventional accelerators). Thus it can be easy to refocus laser-accelerated protons with magnets or other means. The energy spectra with a single-layer metal target are broad and typical of a thermal distribution. We had suggested a method[5] for producing quasi-monoenergetic proton beams using tailored targets. This mechanism has been confirmed with computer simulations[7,43] and has also been experimentally demonstrated.[44] In addition, selective collimation, energy selection (filtering), and focusing even for proton beams with broad energy spectra and broad divergence have been shown to be possible experimentally via an ultra-fast laser-driven microlens.[45]

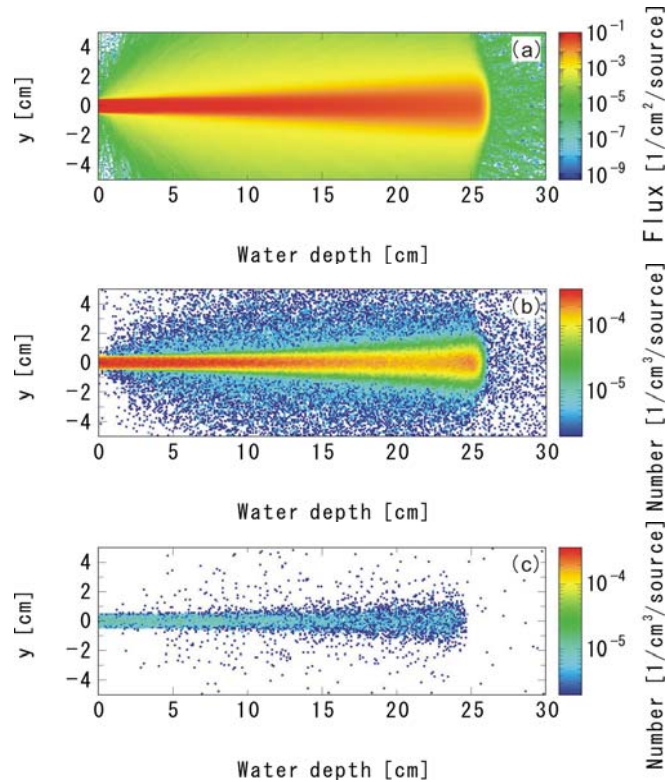

**FIGURE 3**. Monte Carlo simulation of radio-activation of nuclei by proton beams. (a) The flux of protons for the case of a proton pencil beam of energy 200 MeV with a 10 mm diameter irradiating water from the left where the water boundary starts at 5cm. (b) The two-dimensional distribution of residual nuclei: $^{15}$O in water. (c) The two-dimensional distribution of residual nuclei: $^{11}$C in water.

It is also necessary to develop techniques to remove particles other than protons (heavy ions, electrons, gamma-rays, neutrons, etc), which are also emitted from the target as a result of the laser-target interaction.[22] For laser-accelerated protons, the desired beam direction can be set by choosing the laser irradiation direction and magnets are introduced only for the segregation of the desired beam components from other non-desired ones. Therefore, magnetic field levels (and thus the size of magnets) can be almost an order of magnitude lower than those used in a conventional gantry, where a charged beam is typically bent by 90 degrees or 135 degrees.

# OPTIMIZED LASER ACCELERATION

## Methods of investigation

Specific features of laser-accelerated proton therapy require development of simulation tools[46] which include (1) particle-in-cell simulation (PIC) software[47] which calculates the properties of laser-accelerated protons, (2) Monte-Carlo simulation codes PHITS[40] and GEANT4[48] for dose calculations in the human body, and (3) visualization tools for the dose evaluation. We have carried out simulations of laser-accelerated proton therapy using these.

For the Monte-Carlo simulations, CT values, in Hounsfield units, were used to determine a material for each voxel following the work of Schneider.[49] Adjacent CT values were averaged to yield a pixel size of 1.25 by 1.25 mm. The slice thickness was 2.5 mm. A spot scanning method was employed. The spacing between target spots can be specified, or computed automatically based on a pre-computed database of dose profile curves in water. For these tests, we specified a lateral and depth spacing of 2 mm.

Based on earlier PIC simulation results, we chose a Gaussian energy spread of 10% and a beam diameter of 5 mm. A mono-energetic particle beam can be aimed directly at a critical structure, relying on the sharp Bragg peak fall off, but a particle beam with high Gaussian energy spread results in a relatively long distal dose drop-off curve. For this reason, we chose to direct the beam from the side, sometimes even through the bones around the eye socket.

The multi-parametric simulation is a technique, in which a series of many tasks with different sets of the laser and target parameters is performed simultaneously on many processors of a multi-processor supercomputer (Fig. 4). Each set of parameters constitutes an individual task, which is performed using the massively-parallel and fully vectorized code REMP, based on the PIC method and the "density decomposition" scheme.[47] In a two-dimensional model of the laser-plasma interaction, it is possible, in principle, to perform an individual task on a small machine, e. g. personal computer. However, in order to test a large number of different initial conditions with one processor unit it is necessary to run tasks sequentially which can take years to complete. The necessary processing time can be greatly reduced if all the tasks are processed in parallel on a massively-parallel supercomputer or GRID system.

## Target thickness optimization

Characteristics of laser-driven ion beam can be controlled by changing the parameters of the laser pulse and the target. For example, the accelerated ion beam emittance can be set by shaping the foil target – because ions are accelerated in the direction perpendicular to the target surface, bending the foil target can focus the emitted proton beam.[50] The energy spectrum of the ion beam can be set by shaping the target and changing its composition. To obtain quasi-mono-energetic ion beams, the scheme of a double layer target was developed.[5,7,43,51] In this scheme, a sufficiently intense laser pulse sweeps a significant portion of electrons from a thin target building the accelerating electrostatic potential. Figures 5a and 5b show the results of three-dimensional PIC simulation for two cases, when the transverse size of the double-layer target is comparable to the laser focal spot and when it is much greater than the laser focal spot. In both cases targets are irradiated by 30J 65 fs laser pulse with peak intensity $10^{21}$W/cm$^2$. In the first case, the protons are initially localized in a dot coating, with transverse size much less than the laser focal spot, thus ions see an almost homogeneous accelerating field and the resulting energy spectrum is quasi-mono-energetic (Fig. 5a). In the second case, the protons form a wide coating on the rear side of the target. In this case protons, which were initially sufficiently far from the laser pulse axis, acquire less energy than ions near the laser pulse axis, because the transverse scale of the accelerating electric field is comparable to laser focal spot. Although the accumulated spectrum is approximately thermal, the central portion of the ion beam is quasi-mono-energetic since the ion motion is almost laminar. This beamlet can be extracted, e. g., by a screen (Fig. 5b).

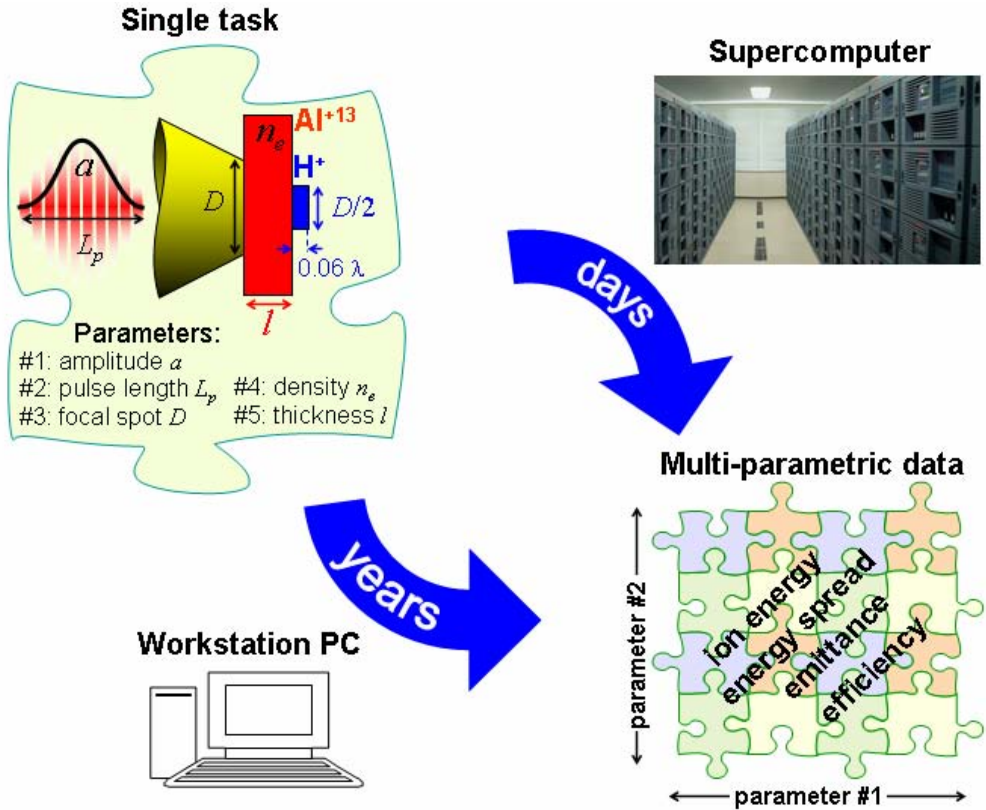

**FIGURE 4.** The scheme of multi-parametric Particle-In-Cell(PIC) simulation. Each individual set of parameters, represented by a tile, is processed in one run by one processor unit. Results from multiple runs inlay a mosaic in the parameter space, from which scaling laws can be seen.

Knowing the dependence of the ion energy on target and laser characteristics is of crucial importance for practical implementations and optimization. Here we present an example of how such the dependence can be foreseen.

We investigate the ion acceleration from a double layer target [5,7,43,52] driven by a laser pulse with multi-parametric Particle-in-Cell (PIC) simulations.[24]

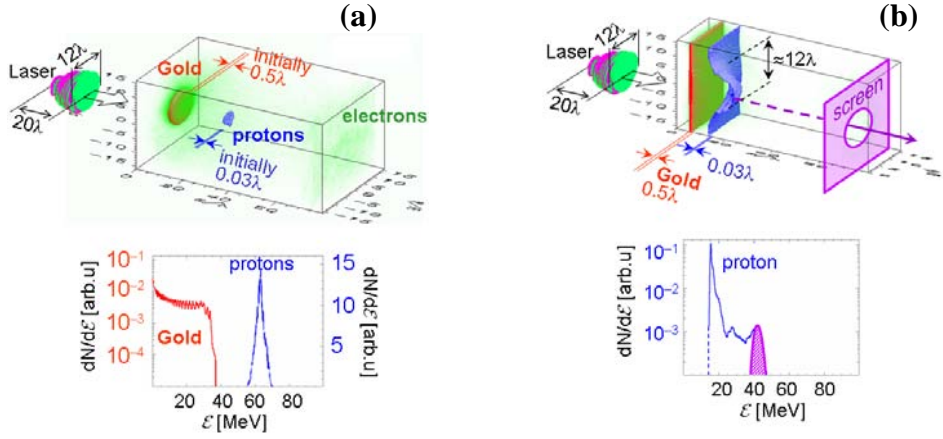

**FIGURE 5.** Toward more mono-energetic beams. (a) Laser-driven ion acceleration at a spot-size double-layer target. (b) Laser-driven ion acceleration at a wide double-layer target.

We analyze the dependence of the interaction products (such as the ion beam energy, emittance, the acceleration time, pulse duration, as well as laser light reflection, transmission and absorption coefficients, etc…) on the following laser pulse and the target parameters: laser pulse intensity $I$, focal spot size $\ell_{las\perp}$ and duration $\ell_{las}/c$, target density $n_e$ and thickness $l_{pl}$. Simulations were performed on 720-processor HP Alpha Server at JAEA-Kansai. In each processor the grid consists of 4016×2176 cells and the number of quasi-particles is $1.9 \times 10^7$. The simulation box size is $251\lambda \times 136\lambda$, where $\lambda$ is the laser wavelength (typically 1 micrometer). The laser pulse initially propagates along the $x$-axis. The target consists of two layers, the first layer is fully stripped aluminum and the second layer is a proton coating. The transverse size of the first layer is $80\lambda$, for the second layer it varies with the laser pulse focal spot size as $\ell_{las\perp}/2$. The second layer is $0.06\lambda$ thick, its density is such that the number of ions in the first layer in the longitudinal direction is $10^3$ times greater than the corresponding number of protons. At chosen conditions, the proton layer is accelerated as a whole. The resulting accelerated ion beam is a pulse of $10^7$-$10^9$ protons with transverse emittance less than 0.3 mm·mrad, (meeting the corresponding requirements of the proton therapy).

We study the idealized model, with a simple target and laser geometry, in order to simplify interpretation and to avoid screening of important material by numerous unimportant details. In experiments, laser pulses have low-intensity intervals, such as pre-pulse (from amplified spontaneous emission (ASE) and other sources), which can alter the structure of the solid target before the main short laser pulse arrives. That is, a preplasma is typically formed, as discussed in, [14][16][42][53]. Despite the idealization, results of our multi-parametric analysis are useful even for interpretation of the laser-foil interaction complicated by the prepulse, which changes the target density and scale length. Upon arrival, the main, high-intensity, part of the laser pulse meets a modified target, which can be approximated as a plasma slab of some density and thickness. From this moment the interaction can be described by our model, because the energy of relativistic processes in the electromagnetic wave interaction with plasmas is significantly higher than that of non-relativistic hydrodynamics associated with melting (evaporating) the thin target. In other words, instead of predicting a particular preplasma condition, our goal is to consider all possible plasma conditions.

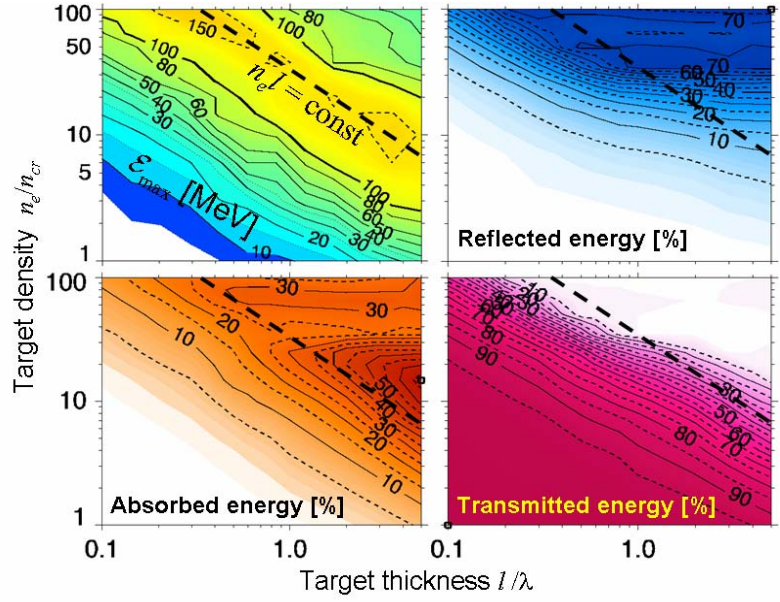

**FIGURE 6.** Optimization of the target thickness for given laser intensity. Dependence of ion beam energy on the target thickness and density, when the target is irradiated by the laser pulse of intensity $5\times10^{21}$ W/cm$^2$, duration 10$\lambda$, and focal spot 10$\lambda$, and a correlation between ion beam energy and laser pulse reflection, absorption and transmission coefficients.

While scanning different target thicknesses $l_{pl}$=0.1…5$\lambda$, densities $n_e$=1…100 $n_{cr}$ (where $n_{cr}=\pi m_e c^2/\lambda^2 e^2$ is the critical density), and laser intensities $I$=10$^{20}$…10$^{22}$ W/cm$^2$ for fixed laser pulse length and focal spot size $\ell_{las}=\ell_{las\perp}$=10$\lambda$, we found that the highest ion energy gain occurs at certain electron areal density of the target $\sigma=n_e l_{pl}$, i. e. the number of electrons integrated along the longitudinal direction per unit area (Fig. 6).

Thus the energy dependence on two parameters $n_e$ and $l_{pl}$ degenerates to a dependence on only one parameter $\sigma$, even though we can identify at least three different mechanisms for the ion acceleration in different regions of parametric space. At small thickness $l_{pl}$ and large density $n_e$ the laser pulse sweeps away a substantial fraction of electrons and the induced strong Coulomb potential of the first layer accelerates protons, in accordance with the scenario described in Refs. [5 7 9 43 52]. At large thickness and small density the laser pulse penetrates through the target and generates a strong quasi-static magnetic field whose pressure causes charge separation which accelerates protons near the plasma-vacuum interface, similar to the mechanism suggested in Refs. [16 54 55]. Somewhere between these two extreme regions we can see the ion acceleration due to ambipolar expansion of the hot electron cloud into vacuum.[56 57 58 59] The degeneracy in dependence of the ion energy on target density and thickness can be interpreted as follows. Ions are accelerated by the electrostatic field of the charge separation caused by the laser pulse. The electrons, if their number is large, can form the electric current which is sufficient to reflect the laser pulse.

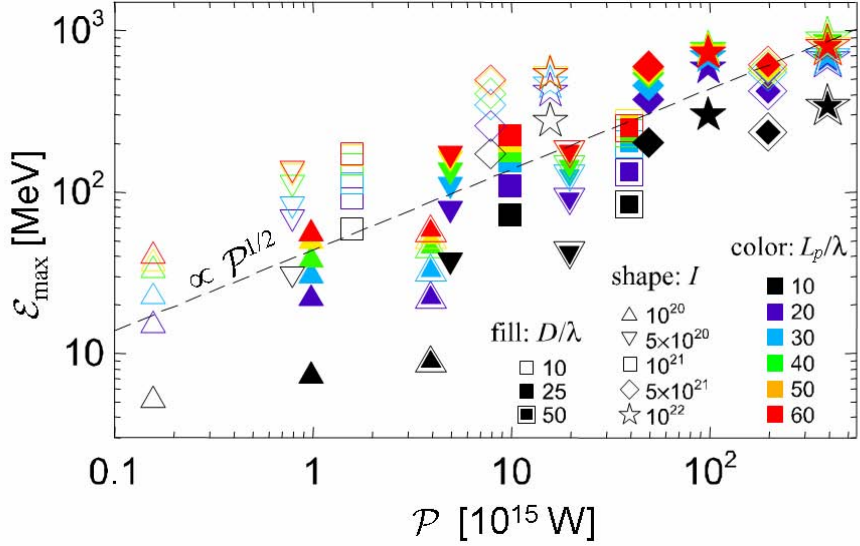

**FIGURE 7.** Dependence of ion beam energy from the laser pulse power.

However this current cannot exceed the limiting electric current $-en_ec$. Therefore, with increasing laser intensity, the reflection becomes less efficient, the plasma becomes more transparent and at some point all the electrons are involved in the interaction. With even greater laser intensity, even more electrons can be disturbed, providing stronger charge separation. Consequently a smaller number of electrons will not be optimal. Thus for every value of the laser intensity there is a certain optimal limiting electric current formed by a certain number of electrons in the slab. Since this effect is mainly one-dimensional, the limiting current is defined by the electron areal density of the slab, as in Refs. [60 61].

The maximum of the ion energy gain corresponds to a particular optimal electron areal density $\sigma_{opt}$. For $\sigma<\sigma_{opt}$, more laser pulse energy is transmitted through the plasma slab. For $\sigma>\sigma_{opt}$, the laser pulse reflection becomes more efficient. At $\sigma=\sigma_{opt}$, the absorption of the laser energy is optimal for the ion acceleration. The proton energy spread is below 5% for $\sigma<\sigma_{opt}$, and it increases for $\sigma>\sigma_{opt}$, as the maximum energy decreases. The optimal electron areal density depends almost linearly on the square root of the laser intensity, $\sigma_{opt}\approx 3+0.4\,I^{1/2}$. This is similar to the criteria of relativistic transparency of a thin foil.[60 61] Varying the $\sigma$, one can find the minimum intensity which gives the desired ion energy gain. In other words, at each intensity there is an optimal $\sigma$ which provides the maximum energy gain. For highest achievable ion energy with each pair of parameters $(l_{pl}, n_e)$, one is led to an optimized scaling

$$\mathcal{E}_p \propto I \qquad (1)$$

Changing the laser pulse duration and the target thickness, we can further control the output ion beam energy. Figure 7 shows the dependence of the energy of the ion beam accelerated with a double-layer target of density $100n_{cr}$ rearranged as a dependence of ion energy on laser pulse power (derived from laser intensity, duration, spot size and target thickness). In the case of thin targets of given thickness and optimal laser pulse duration, the proton energy scales as the square root of the laser pulse power[24]. When the radiation pressure of the laser field becomes dominant, which corresponds to intensities of the order of $10^{22}$ W/cm$^2$, the proton energy becomes proportional to the laser pulse energy and protons can be accelerated to relativistic energies.[62][63] Figure 7 represents the functional dependence with a fixed density. As we shall see in subsection D, if we eliminate the abruptness of the present acceleration, further increase of ion energy is obtainable.

### Energy scaling of ions with respect to the laser pulse length

We discuss the energy of accelerated ions with respect to the laser pulse duration (within the ultra short pulse regime, which is subpicoseconds). The regime presented above for high quality proton beam acceleration requires an adequately high laser pulse intensity. In order to expel almost all the electrons from the focal region the laser electric field, $E_{las}=(4I/c)^{1/2}$, must be larger than the electric field that is formed here due to the electric charge separation

$$E_0 \approx 2\pi n_e e \ell_{pl}. \tag{2}$$

From the geometry of charge formation for sudden blowout sheath acceleration, the longitudinal length of ion acceleration is at most that of the electron acceleration length. This in turn is approximately that of the transverse laser focal spot size $\ell_{las\perp}$. Thus the fast accelerated proton energy is

$$\mathcal{E}_p = eE_0 \ell_{las\perp}. \tag{3}$$

On the other hand, the acceleration length of the electrons is

$$\ell_{acc} = a_0 \lambda \tag{4}$$

where $a_0 = eE_0/m_e\omega c$ is the dimensionless laser pulse amplitude, $\omega$ and $\lambda = 2\pi c/\omega$ are the laser frequency and wavelength respectively. For optimal usage of laser energy this length $\ell_{acc}$ should be also of the order of $\ell_{las}$. In the non-relativistic ion regime of current study, the acceleration time is equal to

$$\tau_{acc} = (2\ell_{las\perp} m_i / eE_0)^{1/2}. \tag{5}$$

until it reaches the plateau value $\ell_{acc} = \ell_{acc}/c$. This is the optimum from both the ion energy and invested laser power. At this optimal point, the required laser pulse length is

$$\ell_{las} = (2c/\omega_{pi})(\ell_{las\perp}/\ell_{pl})^{1/2}. \tag{6}$$

Here $n_{pi}=(4e^2n_e/m_i)^{1/2}$ is the ion plasma frequency.

In order to optimize the proton energy, it is important to increase the laser efficiency and we increase the pulse length and energy for fixed laser power. For example, instead of using ultra-short Ti:sapphire lasers perhaps sub-picoseconds ceramic lasers may serve a purpose because of such scaling. This point may become even more important in the further innovation of adiabatic acceleration discussed in next subsection D.

## Adiabatic acceleration

Since the initial proton acceleration experiment[11] in 2000 at Lawrence Livermore National Laboratory (LLNL), all laser experiments so far have been of the sudden type (and non-adiabatic). In these experiments the focused intense laser irradiates a solid target such as aluminum foil, in which the speed of the electrons reaches the speed of light instantaneously (MeV/μm). In this process the laser energy is first transferred to the electron kinetic energy and then to electrostatic potential energy (and electromagnetic energy due to the large electronic current generated), which in part is converted into the ion kinetic energy. The high density of the target electrons makes the charge separation propagation temporarily stationary followed by acceleration. However, collective acceleration often suffers from the detachment of the laser energy in electrons from that of ions.[64] To overcome this problem in laser ion acceleration, the concept of adiabatic (gradual) change of the accelerating medium was introduced.[65] If and when we can adopt adiabatic[65] (i.e. gradual) acceleration, instead of non-adiabatic (i.e. sudden) acceleration in most (or all) current experiments, we should expect the conversion efficiency from laser energy to charged particle energy to greatly increase. We envision that the adiabaticity of acceleration may be introduced by a variety of techniques not limited by the one originally proposed.[65] For example, Bulanov et al. have already shown the gradual slope of the density can help introduce adiabatic acceleration.[66]

Unlike the sudden sheath blowout scaling discussed above in C, the energy of ions depends on a different set of conditions. The optimal condition of the laser pulse coupling with the plasma slab is introduced, i. e. the condition

$$a_0 = \sigma.$$  (7)

Together with Eq.(4), Eq.(7) yields Eq.(1) once again. This condition is written as

$$a_0 = \frac{\int n d\ell}{n_{cr}\lambda} = \frac{n\ell_{pl}}{n_{cr}\lambda}.$$  (8)

The laser pulse loses a substantial part of its energy after it has propagated over the depletion length[3]

$$\ell_{dep} = \ell_{las}\left(\frac{n_{cr}}{n}\right).$$  (9)

From Eqs. (8) and (9) we find the matching condition between the target length and the laser pulse amplitude:

$$a_0 = \frac{\ell_{las}}{\lambda}.$$  (10)

Here we optimize the thickness of the target to be $\ell_{las} = \ell_{pl}$.

From above written relationships we obtain a scaling for the adiabatically accelerated ion energy in the case of one layer target. It is

$$\mathcal{E}_p = m_e c^2 \left(\frac{\ell_{pl}}{d_e}\right)^2, \tag{11}$$

where $d_e = c/\omega_{pe}$ is the collisionless skin depth, $\omega_{pe} = (4\pi\, n_e e^2/m\gamma)^{1/2}$ is the plasma frequency and $\gamma$ is the Lorentz factor of electrons ($\gamma \sim a_0$ when $a_0 \gg 1$).

In order to provide the matching between the laser pulse and target in the transverse direction, we suggest using the preformed channel with the parabolic profile of the plasma density inside. In this case the matching condition is expressed as

$$\ell_{las} = \sqrt{\lambda R}, \tag{12}$$

where $R$ is the channel radius with $R > l_{pl}$. The optimal situation for the electric field generation occurs when $\ell_{las} = \ell_{pl}$. Note that the plasma thickness is now substantially enhanced to realize adiabatic (gradual) acceleration. On the other hand, we shall see in what follows that the density of the plasma for optimal acceleration is substantially less than the usual solid density. The laser pulse shape also affects its energy depletion length value. A suitably shaped laser pulse loses less energy during interaction with the same profile target.

In order to achieve the energy higher than that given in (11) we consider the graded layer target, in which the relatively high plasma density slab is followed by the inhomogeneous corona with the gradually decreasing density. The ion energy gain can be estimated to be

$$\ell_{las} = \sqrt{\lambda R}, \tag{13}$$

where the electric field is given by $E(x) = en(x)\, \ell_{las}(x)$ with $\ell_{las} = \sqrt{\lambda R}$ and $\mathcal{E}_p = m_i v_i^2/2$. Rewriting Eq. (13) as

$$m_i v_i(x) \frac{dv_i(x)}{dx} = e^2 n(x) \sqrt{\lambda R(x)}, \tag{14}$$

and imposing the condition of adiabatic acceleration,

$$v_i(x) = V(x), \tag{15}$$

where $V(x)$ is the propagation velocity for the accelerating structure, we obtain the equation for the target parameters, $n(x)$ and $R(x)$. The velocity $V(x)$ is determined by the physical mechanism of the electric field generation. The electric field can be produced either by the fast electrons accelerated in the plasma, or by the laser pulse ponderomotive pressure, or by the pressure of quasistatic magnetic field generated by the fast electron electric current.

As an illustration, we consider the case when the velocity $V(x)$ is equal to the laser pulse group velocity $v_g(x)$, which is given by

$$v_g(x) = c \frac{\sqrt{\omega^2 - \omega_{pe}^2(x)}}{\omega}. \quad (16)$$

We note that $v_g(x)$ can be controlled by changing the electron density. We further note that we can make the group velocity equal to zero (or very small) when we need to pick up ions and that we can make it increase as ions gain energy. This is the essence of the innovation of adiabatic acceleration.

Assuming the channel radius, R, to be constant, we find from Eqs. (14)-(16) the exponential profile of the plasma density in the corona to be:

$$n(x) = n_0 \exp\left[-\pi \frac{m_e}{m_i} \frac{x}{\lambda} \left(\frac{R}{\lambda}\right)^{1/2}\right]. \quad (17)$$

For typical parameters, $m_i/m_e \approx 2000$, $R/\lambda \approx 20$, and $\lambda \approx 1\mu m$, we find that the inhomogeneity scale for the plasma corona is of order several hundred microns.

Innovations that are based on insight from the optimizations discussed in this section, affords opportunity to physically realize compact and relatively uncomplicated/uncluttered accelerator systems for beam therapy.

## CLINICAL APPLICATIONS

The combination narrow spot irradiation of laser-driven proton beams coupled with shallow target depth is ideal for an accurate radiotherapy treatment of shallow diseases. A narrow pencil beam of protons will suffer less angular spread due to multiple scattering over the short (several cm) distance to the shallow site of the disease. Such angular broadening increases with increased target depth and limits spatial resolution of the treatment volume.

We are in the early stages of feasibly developing facilities for hadron therapy of oncological diseases. Some key questions arise. Among them are: Can we use a laser driven proton accelerator for clinical practice? What kind of disease will be treated? When can the laser driven accelerator be achieved? Why is the accelerator necessary? Table 2 answers some of these questions. At present, because the laser accelerated proton energy is relatively low, clinical application is limited to diseases existing only a few centimeters under the skin such as small tumors in the eye, thyroid, larynx, nasal or paranasal cavity, breast, superficial lymph node (LN), skin and subcutaneous tissue. One possibility based on HIBMC experiences is to utilize the laser driven accelerator in combination with the standard particle accelerator. This is because the current particle accelerators can generate ion energies high enough to be suitable for deep seated tumors and the laser driven accelerator presently generates lower energy ions.

**TABLE2.** Comparison between a conventional particle accelerator (synchrotron/cyclotron) and a laser proton accelerator (as of the present). As discussed in the text, in the future with a higher-power and better-controlled laser and a better design of the target, it may be possible for laser accelerators to reach energies comparable with conventional accelerators. Rapid development in both of these areas is expected.

|  | Accelerator Systems | |
|---|---|---|
|  | **Synchrotron or Cyclotron** | **Laser ions at present** |
| **Size of machine** | large ($\Phi$30m in HIBMC) | small (table top) |
| **Cost** | Expensive (23 billion yen at HIBMC) | Relatively inexpensive |
| **Energy** | up to 230MeV in HIBMC | several tens of MeV (at present) |
| **Change of energy** | much time | easy and rapid |
| **Range** | 32cm (whole body) | several cm (superficial tumor) |
| **Beam size** | about 10mm | <1mm (at source) |
| **Operation time** | Now available | Petawatt laser JAEA 2001 (experimental) Fox Chase laser 2006 (small size, experimental) GIST laser 2006 (small size, experimental) |
| **Clinical applications** | solid tumor in the body | small sized tumors in the eye, thyroid, larynx, nasal or paranasal cavity, breast, superficial LN, skin, subcutaneous intraoperative or intracavital proton RT |

Below we discuss earlier and easier goals for proton radiotherapy that are derived from the advantageous features as well as today's limitations that characterize laser driven ion acceleration.

In the case of proton therapy of age-related macular degeneration (ARMD), the development of a dedicated machine only for ARMD is indispensable as evidenced by the accurate targeting shown in Figs. 8c and 8d. Because patients develop ARMD in 0.67% (320,000 people) of the population at the age of 50 or over (49,000,000 people) in Japan, and 1.7 % (830,000 people) a year in a more recent report,[67] many patients cannot be treated in current conventional accelerator facilities.

Simulation studies have shown that laser accelerated protons and particle selection systems have the capability to deliver superior radiation therapy treatment.[31] With our PIC and Monte-Carlo simulation, we demonstrated treatment planning for eye diseases which may be treated using the spot-scanning technique by laser ion accelerated proton beams.[46] Using PIC simulations the possible available proton beam energies with currently available lasers were determined. Based on these results various input parameters were chosen for the Monte-Carlo simulations with GEANT4. We show the results for the cases: Uveal melanoma and ARMD. In both cases the energies are somewhat higher due to the propagation of the beam through bone tissue.

Uveal melanoma (Figs. 8a and 8b)[68 69]

The proton radiotherapy for uveal melanomas was simulated. Minute proton beams delivered a range of the kinetic energy to the target spots. A right oblique portal was adopted to minimize exposure to the normal structures of the eye. The energy range was from 38 to 81 MeV using a Gaussian beam energy spread of 10%. Proton therapy may be used for patients who are not suitable for more conservative therapies.[69]

ARMD (Figs. 8c and 8d)

Left lateral proton irradiation was planned. The clinical target volume was settled on the macula of the left eye. The iso-dose curve shows the advantage of proton radiotherapy. The parameters are the same as in the previous case, but the energy range was from 61 to 75 MeV.

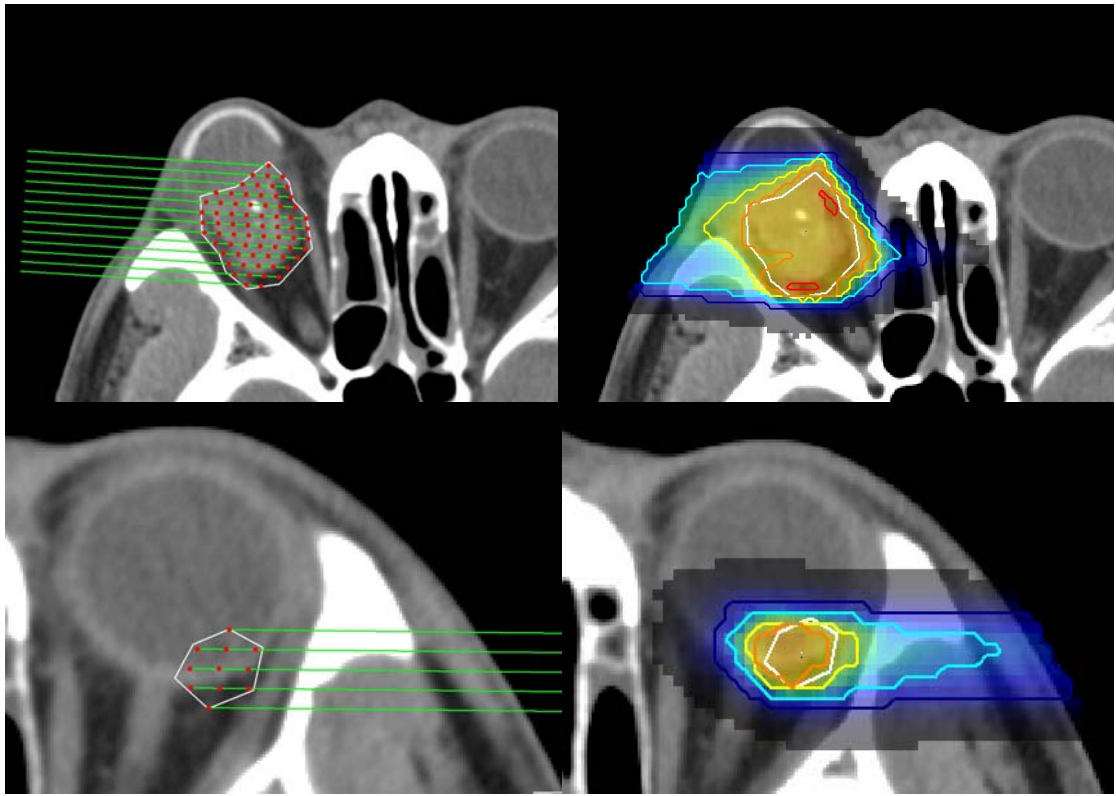

**FIGURE 8.** Monte Carlo simulation of the spot-scanning simulation of laser proton radiotherapy for eye melanoma (a,b) and ARMD (c,d). Particle-in-cell simulation (PIC) results are used to determine the properties of laser-accelerated protons. Based on these results parameters are then chosen for the input into the Monte-Carlo simulation code and visualization tools for the dose evaluation are used to image the results. The initial input parameters are determined by forward planning. After the Monte-Carlo simulation, the particle counts were fine tuned to achieve a smooth dose distribution. Curves correspond to 25%, 50%, 75%, 90% and 110% of the required dose.

# CONCLUSIONS

The laser proton accelerator concept has characteristics that are compact and enables new ways to practice radio-therapy. A laser driven proton accelerator can facilitate changing the proton energy easily and rapidly such that fine scanning irradiation using a minute spot becomes easily realized as demonstrated in our Monte-Carlo dose distribution simulation. We have shown that the compactness of the system makes more readily available *in situ* real-time diagnostics such as PET for realizing BCRT. The optimization techniques for proton energy maximization and a new acceleration method (adiabatic acceleration) introduced in the present work contribute to greatly increase the achievable ion energy for a given laser intensity and to enhance the conversion efficiency of laser energy to charged particle energy. Presently available laser accelerated proton sources may be used for treating shallow tumors and ARMD as an earlier adoption of the laser acceleration technique to medical applications. We should not underestimate technical challenges lying ahead for clinical usage starting from the currently available parameters, while we proceed to resolve each of these issues with the consorted efforts of the medical, physical, and computational communities. Radically different from the conventional accelerator, the laser ion accelerator is small and suitable for a minute target and has great potential to improve the availability and the method of ion therapy with a series of innovations in the future. It is important to further intensify research in the field for the advancement of the current technique.

# ACKNOWLEDGMENTS

This work was supported by JST-CREST and the Special Coordination Fund for Promoting Science and Technology of MEXT, Japan. The authors are grateful for support provided by A. Fushio, Y. Miyazaki, K. Fukuhara, and K. Yokoi.

# REFERENCES

1. M. Goitein, A.J. Lomax, E.S. Pedroni, "Treating cancer with protons," *Phys. Today.* **55**, 45-50 (2002).
2. J. Sisterson, "A newsletter for those interested in proton, light ion and heavy charged particle radiotherapy," *Particles* **36**, 1-11 (2005)
3. T. Tajima and J. M. Dawson, "Laser electron accelerator," *Phys. Rev. Lett.* **43**, 267-270 (1979).
4. T. Tajima, "Prospect for compact medical laser accelerators [abstract]," *J. Jpn. Soc. Ther. Radiol. Oncol.* **9 (Suppl 2)**, 83 (1997).
5. S.V. Bulanov and V.S. Khoroshkov, "Feasibility of using laser ion accelerators in proton therapy," *Plasma Phys. Rep.* **28**, 453-456 (2002).
6. E. Fourkal, B. Shahine, M. Ding, J.S. Li, T. Tajima, and C.-M. Ma, "Particle in cell simulation of laser-accelerated proton beams for radiation therapy," *Med. Phys.* **29**, 2788-2798 (2002).
7. S.V. Bulanov, T.Zh. Esirkepov, V.S. Khoroshkov, A.V. Kuznetsov, and F. Pegoraro, "Oncological hadrontherapy with laser ion accelerators," *Phys. Lett. A* **299**, 240-247 (2002).

8. V. Malka, S. Fritzler, E. Lefebvre, E. d'Humières, R. Ferrand, G. Grillon, C. Albaret, S. Meyroneinc, J.P. Chambaret, A. Antonetti, and D. Hulin, "Practicability of protontherapy using compact laser systems," *Med. Phys.* **31**, 1587-1592 (2004).
9. S.V. Bulanov, H. Daido, T.Zh. Esirkepov, V.S. Khoroshkov, J. Koga, K. Nishihara, F. Pegoraro, T. Tajima, and M. Yamagiwa, "Feasibility of using laser ion accelerators in proton therapy," <u>In the Physics of Ionized Gases,</u> *AIP Conference Proceedings* **740**, 414-429 (2004).
10. Y. Glinec, J. Faure, V. Malka, T. Fuchs, H. Szymanowski, and U. Oelfke, "Radiotherapy with laser-plasma accelerators: Monte Carlo simulation of dose deposited by an experimental quasimonoenergetic electron beam," *Med. Phys.* **33**, 155-162 (2006).
11. R.A. Snavely, M.H. Key, S.P. Hatchett, T.E. Cowan, M. Roth, T.W. Phillips, M.A. Stoyer, E.A. Henry, T.C. Sangster, M.S. Singh, S.C. Wilks, A. MacKinnon, A. Offenberger, D.M. Pennington, K. Yasuike, A.B. Langdon, B.F. Lasinski, J. Johnson, M.D. Perry, and E.M. Campbell, "Intense high-energy proton beams from Petawatt-laser irradiation of solids," *Phys. Rev. Lett.* **85**, 2945-2948 (2000).
12. E.L. Clark, K. Krushelnick, J.R. Davies, M. Zepf, M. Tatarakis, F.N. Beg, A. Machacek, P.A. Norreys, M.I.K. Santala, I. Watts, and A.E. Dangor, "Measurements of energetic proton transport through magnetized plasma from intense laser interactions with solids," *Phys. Rev. Lett.* **84**, 670-673 (2000).
13. A. Maksimchuk, S. Gu, K. Flippo, D. Umstadter, and V.Yu. Bychenkov, "Forward ion acceleration in thin films driven by a high-intensity laser," *Phys. Rev. Lett.* **84**, 4108-4111 (2000).
14. A.J. MacKinnon, M. Borghesi, S. Hatchett, M.H. Key, P.K. Patel, H. Campbell, A. Schiavi, R. Snavely, S.C. Wilks, and O. Willi, "Effect of plasma scale length on multi-MeV proton production by intense laser pulses," *Phys. Rev. Lett.* **86**, 1769-1772 (2001).
15. A.J. Mackinnon, Y. Sentoku, P.K. Patel, D.W. Price, S. Hatchett, M.H. Key, C. Andersen, R. Snavely, and R.R. Freeman, "Enhancement of proton acceleration by hot-electron recirculation in thin foils irradiated by ultraintense laser pulses," *Phys. Rev. Lett.* **88**, 215006 (2002).
16. K. Matsukado, T. Esirkepov, K. Kinoshita, H. Daido, T. Utsumi, Z. Li, A. Fukumi, Y. Hayashi, S. Orimo, M. Nishiuchi, S.V. Bulanov, T. Tajima, A. Noda, Y. Iwashita, T. Shirai, T. Takeuchi, S. Nakamura, A. Yamazaki, M. Ikegami, T. Mihara, A. Morita, M. Uesaka, K. Yoshii, T. Watanabe, T. Hosokai, A. Zhidkov, A. Ogata, Y. Wada, and T. Kubota, "Energetic protons from a few-micron metallic foil evaporated by an intense laser pulse," *Phys. Rev. Lett.* **91**, 215001 (2003).
17. D. Umstadter, "Relativistic laser–plasma interactions," *J. Phys. D.* **36**, R151-R165 (2003).
18. G.A. Mourou, T. Tajima, S.V. Bulanov, "Optics in the relativistic regime," *Rev. Mod. Phys.* **78**, 309-371 (2006).
19. M. Borghesi, J. Fuchs, S.V. Bulanov, A.J. MacKinnon, P.K. Patel, and M. Roth, "Fast ion generation by high-intensity laser irradiation of solid targets and applications," *Fus. Sc. Techn.* **49**, 412-439 (2006).
20. W. H. Scharf, *Biomedical Particle Accelerators* (AIP Press, New York, 1994).
21. A. Fukumi, M. Nishiuchi, H. Daido, Z. Li, A Sagisaka, K. Ogura, S. Orimo, M. Kado, Y. Hayashi, M. Mori, S.V. Bulanov, and T. Esirkepov, "Laser polarization dependence of proton emission from a thin foil target irradiated by a 70 fs, intense laser pulse," *Phys. Plasmas* **12**, 100701 (2005).
22. E. Fourkal, J.S. Li, M. Ding, T. Tajima, and C.-M. Ma, "Particle selection for laser-accelerated proton therapy feasibility study," *Med. Phys.* **30**, 1660-1670 (2003).
23. A. W. Chao and M. Tigner, eds., *Handbook of Accelerator Physics and Engineering* (World Scientific, 1999).
24. T. Esirkepov, M. Yamagiwa, and T. Tajima, "Laser ion-acceleration scaling laws seen in multiparametric particle-in-cell simulations," *Phys. Rev. Lett.* **96**, 105001 (2006).
25. W. Enghardt, P. Crespo, F. Fielder, R. Hinz, K. Parodi, J. Pawelke, and F. Pönisch, "Charged hadron tumour therapy monitoring by means of PET," *Nucl. Instrum. Meth. Phys. Res. A* **525**, 284-288 (2004).
26. L.Xing, B. Thorndyke, E. Schreibmann, Y. Yang, T-F. Li, G-Y. Kim, G. Luxton, and A. Koong, "Overview of image-guided radiation therapy," *Medical Dosimetry* **31**, 91-112 (2006).
27. W.T. Chu, B.A. Ludewigt, and T.R. Renner, "Instrumentation for treatment of cancer using proton and light-ion beams," *Rev. Sci. Instrum.* **64**, 2055-2122 (1993).

28. T. Furukawa, K. Noda, T.H. Uesugi, T. Naruse, and S. Shibuya, "Intensity control in RF-knockout extraction for scanning irradiation," *Nucl. Instrum. Meth. B* **240**, 32-35 (2005).
29. T. E. Cowan, J. Fuchs, H. Ruhl, A. Kemp, P. Audebert, M. Roth, R. Stephens, I. Barton, A. Blazevic, E. Brambrink, J. Cobble, J. Fernández, J.-C. Gauthier, M. Geissel, M. Hegelich, J. Kaae, S. Karsch, G. P. Le Sage, S. Letzring, M. Manclossi, S. Meyroneinc, A. Newkirk, H. Pépin, and N. Renard-LeGalloudec, "Ultralow Emittance, Multi-MeV Proton Beams from a Laser Virtual-Cathode Plasma Accelerator". *Phys. Rev. Lett.* **92**, 204801 (2004).
30. A.K. Carlsson, P. Andreo, and A. Brahme, "Monte Carlo and analytical calculation of proton pencil beams for computerized treatment plan optimization," *Phys. Med. Biol.* **42**, 1033-1053 (1997).
31. E. Fourkal, J. S. Li, W. Xiong, A. Nahum and C-M. Ma, "Intensity modulated radiation therapy using laser-accelerated protons: a Monte Carlo dosimetric study," *Phys. Med. Biol.* **48**, 3977-4000 (2003).
32. A. W. Beavis, "Is tomotherapy the future of IMRT?," *The British Journal of Radiology* **77**, 285-295 (2004).
33. Y. Hishikawa, K. Kagawa, M. Murakami, H. Sakai, T. Akagi, and M. Abe, "Usefulness of positron-emission tomographic images after proton therapy," *Int. J. Radiat. Oncol. Biol. Phys*. **53**, 1388-1391 (2002).
34. T. Nishio, T. Sato, H. Kitamura, K. Murakami, and T. Ogino, "Distributions of β+ decayed nuclei generated in the CH2 and H2O targets by the target nuclear fragment reaction using therapeutic MONO and SOBP proton beam," *Med. Phys.* **32**, 1070-1082 (2005).
35. E. Urabe, T. Kanai, M. Kanazawa, A. Kitagawa, K. Noda, T. Tomitani, M. Suda, Y. Iseki, K. Hanawa, K. Sato, M. Shimbo, H. Mizuno, Y. Hirata, Y. Futami, Y. Iwashita, and A. Noda, "Spot scanning using radioactive 11C beams for heavy-ion radiotherapy," *Jpn J. Appl. Phys.* **40**, 2540-2548 (2001).
36. K. Parodi, P. Crespo, H. Eickhoff, T. Haberer, J. Pawelke, D. Schardt, and W. Enghardt, "Random coincidences during in-beam PET measurements at microbunched therapeutic ion beams," *Nucl. Instrum. Meth. Phys. Res. A* **545**, 446-458 (2005).
37. P. Crespo, T. Barthel, H. Frais-Kolbl, E. Griesmayer, K. Heidel, K. Parodi, J.P. Pawelke, and W. Enghardt, "Suppression of random coincidences during in-beam PET measurements at ion beam radiotherapy facilities," *IEEE Trans. Nucl. Sci.* **52**, 980-987 (2005).
38. P. Crespo, G. Shakirin and W. Enghardt, "On the detector arrangement for in-beam PET for hadron therapy monitoring," *Phys. Med. Biol.* **51**, 2143-2163 (2006).
39. K. Parodi, W. Enghardt and T. Haberer, "In-beam PET measurements of $\beta^+$ radioactivity induced by proton beams," *Phys. Med. Biol.* **47**, 21-36 (2002).
40. H. Iwase, K. Niita, and T. Nakamura, "Development of general-purpose particle and heavy ion transport Monte Carlo code," *J. Nucl. Sci. Tech.* **39**, 1142-1151 (2002).
41. M.B. Chadwick, D.T.L. Jones, G.J. Arendse, A.A. Cowley, W.A. Richter, J.J. Lawrie, R.T. Newman, J.V. Pilcher, F.D. Smit, G.F. Steyn, JW Koen and JA Stander, "Nuclear interaction cross sections for proton radiotherapy," *Nuclear Physics A* **654**, 1051c-1057c (1999).
42. Y. Sentoku, V.Y. Bychenkov, K. Flippo, A. Maksimchuk, K. Mima, G. Mourou, Z.M. Sheng, and D. Umstadter, "High-energy ion generation in interaction of short laser pulse with high-density plasma," *Appl. Phys. B* **74**, 207-215 (2002).
43. T.Zh. Esirkepov, S.V. Bulanov, K. Nishihara, T. Tajima, F. Pegoraro, V.S. Khoroshkov, K. Mima, H. Daido, Y. Kato, Y. Kitagawa, K. Nagai, and S. Sakabe, "Proposed double-layer target for the generation of high-quality laser-accelerated ion beams," *Phys. Rev. Lett.* **89**, 175003 (2002).
44. H. Schwoerer, S. Pfotenhauer, O. Jäckel, K.-U. Amthor, B. Liesfeld, W. Ziegler, R. Sauerbrey, K.W.D. Ledingham, and T. Esirkepov, "Laser-plasma acceleration of quasi-monoenergetic protons from microstructured targets," *Nature* **439**, 445-448 (2006).
45. T. Toncian, M. Borghesi, J. Fuchs, E. d'Humières, P. Antici, P. Audebert, E. Brambrink, C.A. Cecchetti, A. Pipahl, L. Romagnani, and O. Willi, "Ultrafast Laser-Driven Microlens to Focus and Energy-Select Mega-Electron Volt Protons," *Science* **312**, 410-413 (2006).
46. M. Murakami, S. Miyajima, K. Sutherland, S. Bulanov, T. Esirkepov, J. Koga, K. Yamaji, M. Yamagiwa, Y. Hishikawa, and T. Tajima, "Possibility of laser-accelerated proton beams in radiotherapy [abstract]," *E. J. C.* (**Supple 3**), 410 (2005).

47. T. Esirkepov, "Exact charge conservation scheme for Particle-in-Cell simulation with an arbitrary form-factor," *Computer Physics Communications* **135**, 144-153 (2001).
48. S. Agostinelli, et al., "GEANT4-a simulation toolkit," *Nucl. Instrum. Methods Phys. Res. A* **506**, 250-303 (2003).
49. W. Schneider, T. Bortfeld, and W. Schlegel, "Correlation between CT numbers and tissue parameters needed for Monte Carlo simulations of clinical dose distributions," *Phys. Med. Biol.* **45**, 459-478 (2000).
50. M. Roth, T. E. Cowan, M. H. Key, S. P. Hatchett, C. Brown, W. Fountain, J. Johnson, D. M. Pennington, R. A. Snavely, S. C. Wilks, K. Yasuike, H. Ruhl, F. Pegoraro, S. V. Bulanov, E. M. Campbell, M. D. Perry, and H. Powell, "Fast Ignition by Intense Laser-Accelerated Proton Beams". *Phys. Rev. Lett.* **86**, 000436 (2001).
51. S. V. Bulanov, T. Zh. Esirkepov, F. F. Kamenets, Y. Kato, A. V. Kuznetsov, K. Nishihara, F. Pegoraro, T. Tajima, and V. S. Khoroshkov," Generation of high-quality charged particle beams during the acceleration of ions by highpower laser radiation," *Plasma Physics Reports.* **28**, 975-991 (2002);
52. S. V. Bulanov, T. Zh. Esirkepov, F. F. Kamenets, Y. Kato, A. V. Kuznetsov, K. Nishihara, F. Pegoraro, T. Tajima, and V. S. Khoroshkov, "Generation of high-quality charged particle beams during the acceleration of ions by highpower laser radiation," *Plasma Physics Reports.* **28**, 975-991 (2002).
53. K. Nemoto, A. Maksimchuk, S. Banerjee, K. Flippo, G. Mourou, D. Umstadter, and V. Yu. Bychenkov, "Laser-triggered ion acceleration and table top isotope production," *Appl. Phys. Lett.* **78**, 595-597 (2001).
54. A. V. Kuznetsov, T. Zh. Esirkepov, F. F. Kamenets, and S. V. Bulanov, "Efficiency of ion acceleration by a relativistically strong laser pulse in an underdense plasma," *Plasma Phys. Rep.* **27**, 211-220 (2001).
55. H. Amitani, T. Esirkepov, S. Bulanov, K. Nishihara, A. Kuznetsov, and F. Kamenets, "Accelerated dense ion filament formed by ultra intense laser in plasma slab," *AIP Conference Proceedings* **611**, 340-345 (2002).
56. A. V. Gurevich, L. V. Pariskaya, and L. P. Pitaievskii, "Self-similar motion of rarefied plasma," *Sov. Phys. JETP* **22**, 449-454 (1966).
57. S. C. Wilks, A. B. Langdon, T. E. Cowan, M. Roth, M. Singh, S. Hatchett, M. H. Key, D. Pennington, A. MacKinnon, and R. A. Snavely, "Energetic proton generation in ultra-intense laser–solid interactions," *Phys. Plasmas* **8**, 542-549 (2001).
58. P. Mora, "Plasma Expansion into a Vacuum," *Phys. Rev. Lett.* **90**, 185002 (2003).
59. S. V. Bulanov, T. Zh. Esirkepov, J. Koga, T. Tajima, and D. Farina, "Concerning the maximum energy of ions accelerated at the front of a relativistic electron cloud expanding into vacuum," *Plasma Phys. Rep.* **30**, 18-29 (2004).
60. V. A. Vshivkov, N. M. Naumova, F. Pegoraro, and S. V. Bulanov, "Nonlinear electrodynamics of the interaction of ultra-intense laser pulses with a thin foil," *Phys. Plasmas* **5**, 2727-2741 (1998).
61. S. V. Bulanov, F. Califano, G. I. Dudnikova, T. Z. Esirkepov, I. N. Inovenkov, F. F. Kamenets, T. V. Lisejkina, M. Lontano, K. Mima, N. M. Naumova, K. Nishihara, F. Pegoraro, H. Ruhl, A. S. Sakharov, Y. Sentoku, V.A.Vshivkov, and V. V. Zhakhovskii, "Relativistic interaction of laser pulses with plasmas," in *Reviews of Plasma Physics*, Vol. **22**, edited by V. D. Shafranov (Kluwer Academic/Plenum, New York, 2001) pp. 227-335.
62. T. Esirkepov, M. Borghesi, S. V. Bulanov, G. Mourou, and T. Tajima, "Highly Efficient Relativistic-Ion Generation in the Laser-Piston Regime," *Phys. Rev. Lett.* **92**, 175003 (2004).
63. S. V. Bulanov, T. Zh. Esirkepov, J. Koga, and T. Tajima, "Interaction of electromagnetic waves with plasma in the radiation-dominated regime," *Plasma Phys. Rep.* **30**, 196-213 (2004).
64. F. Mako and T. Tajima, "Collective ion acceleration by a reflexing electron beam: Model and scaling," *Phys. Fluids* **27**, 1815-1820 (1984).
65. B. Rau and T. Tajima, "Strongly nonlinear magnetosonic waves and ion acceleration," *Phys. Plasmas* **5**, 3575-3580 (1998).
66. S.V. Bulanov, D.V. Dylov, T.Zh. Esirkepov, F.F. Kamenets, and D.V. Sokolov, "Ion acceleration in a dipole vortex in a laser plasma corona," *Plasma. Phys. Rep.* **31**, 369-381 (2005).

67. M. Miyazaki, Y. Kiyohara, A. Yoshida, M. Iida, Y. Nose, and T. Ishibashi, "The 5-year incidence and risk factors for age-related maculopathy in a general Japanese population: the Hisayama study," *Invest. Ophthalmol. Vis. Sci.* **46**, 1907-1910 (2005).
68. M.F. Moyers, R.A. Galindo, L.T. Yonemoto, L. Lore, E.J. Friedrichsen, M.A. Kirby, J.D. Slater, and J.M. Slater, "Treatment of macular degeneration with proton beams," *Med. Phys.* **26**, 777-782 (1999).
69. B. Damato, A. Kacperek, M. Chopra, I.R. Campbell, and R.D. Errington, "Proton beam radiotherapy of choroidal melanoma: The Liverpool-Clatterbridge experience," *Int. J. Radiat. Oncol. Biol. Phys.* **62**, 1405-1411 (2005).